\documentclass[aps,prd,twocolumn,showpacs,showkeys,nofootinbib]{revtex4}
\UseRawInputEncoding
\usepackage{epsfig,graphics,color}
\usepackage{hyperref}
\usepackage{subfigure}
\usepackage{mathtools}
\usepackage{amssymb}
\usepackage{amsmath}
\usepackage[normalem]{ulem}

\newlength{\llslash}

\newcommand{\tr}{\rm tr \,}

\newcommand{\CD}{\hat{\partial}}

\begin{document}

\title{Chiral excitations of open-beauty systems }
\author{Xiao-Yu Guo$^{1,2}$, Matthias F.M. Lutz$^{2,3}$}
\affiliation{$^1$ Institute of Theoretical Physics, Faculty of Science,\\
Beijing University of Technology, Beijing 100124, China}
\affiliation{$^2$ GSI Helmholtzzentrum f\"ur Schwerionenforschung GmbH,\\
Planck Str. 1, 64291 Darmstadt, Germany}
\affiliation{$^3$ Technische Universit\"at Darmstadt, D-64289 Darmstadt, Germany}

\begin{abstract}
We study the  scattering of open-beauty mesons  and Goldstone bosons as predicted by the chiral SU(3) Lagrangian.
The impact of subleading-order chiral interactions to  systems with $J^P= 0^+$ and $J^P=1^+$ quantum numbers is worked out. We estimate the relevant low-energy coefficients from the open-charm sector, for which their values have been determined previously from sets of QCD lattice data. The leading-order heavy-quark symmetry-breaking effects are estimated by matching the $B$-meson ground-state chiral mass formula to the mass formula from the heavy-quark effective theory. We make refined predictions for the flavor antitriplet and sextet resonances that are generated dynamically by coupled-channel interactions.

\end{abstract}

\maketitle

\section{Introduction}

How to understand the spectrum of QCD is the long-lasting challenge in contemporary theoretical physics.
While the strong interactions lead to a myriad of hadronic resonances,
QCD, the fundamental theory, fails to describe these phenomena if tackled with a perturbative expansion in the gauge coupling constant. 
In order to unravel nonperturbative aspects of QCD, effective field theories are commonly employed. They are constructed to respect the asymptotic symmetries of QCD in its low-energy limit.
Those symmetries include the chiral SU(3) symmetry (which holds exactly when the quarks $u$, $d$, and $s$ are massless) and the heavy-quark symmetry (which asymptotically holds as the quarks $c$ or $b$ tend to be infinitely massive). 
An open-charm or open-beauty system, which is composed of one heavy quark $c$ or $b$ and light quarks, possesses both of these two symmetries and therefore plays a crucial role in studies of QCD.

Chiral SU(3) symmetry drives the interactions between an open heavy-flavor meson and Goldstone bosons. Such interactions can be described by an effective chiral Lagrangian. The leading order s-wave interaction may be attractive or repulsive and resonant states can thus be dynamically generated (see e.g. \cite{Lutz:2001yb}). According to such an interaction, the nature of the open-charm meson $D_{s0}^*(2317)$ can be successfully explained.
While within a quark model approach  \cite{DiPierro:2001dwf,Albaladejo:2018mhb} such state poses a puzzle, coupled-channel scattering studies  between Goldstone bosons and $J^P = 0^-$ $D$-meson ground states suggest that it is a bound state below the  $DK$ threshold \cite{Kolomeitsev:2003ac,Guo:2006fu,Lang:2014yfa}.
It comes as a part of a flavor antitriplet, which is completed by a broad state with isospin-strangeness $(I,S) = (1/2,0)$ quantum numbers. 
The heavy-quark spin symmetry implies that a heavy-flavor hadron always comes 
with almost degenerate spin partners. This reflects the marginal importance of the spin orientation of its heavy-quark content.  
For instance, the $D_{s1}(2460)$ can be understood as the spin 1 partner of the $D_{s0}^*(2317)$. In addition, the heavy-quark flavor symmetry suggests that open-beauty partners of the open-charm states should exist. We note, however, that so far 
no experimental evidence for the open-beauty partner of the $D_{s0}^*(2317)$  has 
been found. 
How stable are these theoretical predictions? Answering this question requires 
systematic studies that take into account  chiral and heavy-quark symmetry-breaking effects as based on effective Lagrangians suitably linked to QCD. 
In our work we use chiral perturbation theory (ChPT) and heavy-quark effective theory (HQET). 

Chiral symmetry-breaking effects are encoded into low-energy constants (LEC) of the chiral Lagrangian \cite{Lutz:2007sk,Guo:2008gp,Liu:2012zya,Altenbuchinger:2013vwa,Guo:2018tjx}. 
In a recent study \cite{Guo:2018kno}, the LECs relevant in the open-charm chiral Lagrangian have been well estimated from QCD lattice data on ensembles 
with a variety of unphysical quark masses. It was shown that the open-charm mesons $D_{s0}^*(2317)$ and $D_{s1}(2460)$ are stable against the next-to-leading (NLO) chiral corrections \cite{Guo:2018kno,Guo:2018gyd}. 
Moreover, the same work \cite{Guo:2018kno} predicted the behavior of the $(1/2,0)$ antitriplet partner of the $D_{s0}^*(2317)$ at unphysical quark masses, which has been recently confirmed by lattice simulations \cite{Gayer:2021xzv}.

Motivated by our success in the open-charm sector, we would like to derive more 
quantitative predictions for the open-beauty partner systems.  
In this article, we will study the chiral symmetry-breaking effects in the open-beauty systems, working out systematically heavy-quark spin and flavor symmetry-breaking effects. Such a projection requests a matching of ChPT to HQET. 

We will first recall the part of the chiral Lagrangian responsible for the leading-order symmetry-breaking LECs. The heavy-quark mass dependence of those LECs is determined by matching the chiral formula of the $B$-meson masses to the HQET result. Based on such LECs, we will derive the pole positions of resonance states in the complex plane as generated by coupled-channel scattering from the chiral Lagrangian.
Special attention will be paid to the antitriplet states. Predictions for the open-beauty partners of $D_{s0}^*(2317)$ and $D_{s1}(2460)$ will be made.
A brief discussion on the flavor-exotic sextet states will follow \cite{Kolomeitsev:2003ac,Hofmann:2003je}. In particular, we will comment on the disputed isospin and strangeness one  $X(5568)$ state \cite{D0:2016mwd,Aaij:2016iev,Aaltonen:2017voc,Sirunyan:2017ofq,Aaboud:2018hgx,Agaev:2016mjb,Lang:2016jpk,Burns:2016gvy,Lu:2016kxm,Chen:2016mqt,Liu:2016xly,Yang:2016sws,Ke:2018stp}.

\section{The chiral Lagrangian with heavy-quark flavor symmetry}

The pseudoscalar $B$-meson ground states form a flavor antitriplet $B = (B^0,-B^+,B_s^+)$. They are the heavy-quark flavor partners of the open-charm antitriplets $D = (D^0,-D^+,D_s^+)$ states. In both flavor sectors the heavy-quark spin symmetry predicts almost degenerate $J^P = 1^-$ partner states. According to the heavy-quark flavor symmetry, the chiral Lagrangian for these two types of mesons take the same form.  Therefore it is useful to introduce 
generic heavy-meson fields $H$ and $H_{\mu \nu}$, that refer either to a $D$ or a $B$-meson antitriplet field. The kinetic terms of the heavy-flavored mesons constitute the leading-order chiral Lagrangian
\begin{eqnarray}
&& \mathcal{L}_{\mathrm{kin}}=(\CD_\mu H)(\CD^\mu \bar H)- \Big(\bar M - \frac{3}{4}\Delta\Big)^2\, H \, \bar H \nonumber\\
&& \ \ -(\CD_\mu H^{\mu\alpha})(\CD^\nu \bar H_{\nu\alpha})
 +\,\frac{1}{2} \Big(\bar M +\frac{1}{4}\Delta\Big)^2 H^{\mu\alpha} \,\bar H_{\mu\alpha} ,
\label{def-kin}
\end{eqnarray}
where following \cite{Guo:2018kno,Guo:2018gyd} the tensor field  representation is adopted for the vector mesons.
The LECs  $\bar M - \frac{3}{4}\Delta$ and $\bar M +\frac{1}{4}\Delta$ are the chiral-limit masses of the heavy pseudoscalar and vector mesons respectively. They depend on the heavy-quark mass $M_Q$. 
The $\bar M=\bar M(M_Q)$ scales as $\bar M\sim M_Q$ in the heavy-quark mass limit. The hyperfine splitting $\Delta = \Delta(M_Q)$ is  caused by $O(1/M_Q)$ heavy-quark spin symmetry-breaking effect \cite{Wise:1992hn,Goity:1992tp}.

The chiral covariant derivative $ \CD_\mu$ in (\ref{def-kin}) 
\begin{eqnarray}
&& \CD_\mu \bar H = \partial_\mu \, \bar H + \Gamma_\mu\,\bar H \,,  \qquad \;\; 
\CD_\mu H = \partial_\mu \,H  - H\,\Gamma_\mu \,.
\nonumber\\    
&& \Gamma_\mu ={\textstyle \frac{1}{2}}\,e^{-i\,\frac{\Phi}{2\,f}} \,\partial_\mu  \,e^{+i\,\frac{\Phi}{2\,f}}
+{\textstyle \frac{1}{2}}\, e^{+i\,\frac{\Phi}{2\,f}} \,\partial_\mu \,e^{-i\,\frac{\Phi}{2\,f}},
\label{def-derivative}
\end{eqnarray}
involves the flavor octet Goldstone-boson fields as encoded in the $3\times 3$ matrix $\Phi$. 
The parameter $f$ is the chiral limit of the pion-decay constant for which we choose $f = 92.4$ MeV. The leading-order, so called Tomozawa-Weinberg, interaction between the $B$ mesons and the Goldstone bosons is implied by the kinetic terms (\ref{def-kin}) via its covariant derivative and the chiral connection $\Gamma_\mu$ .

At the NLO the chiral Lagrangian has terms that are proportional to the masses of the  $u$-, $d$-, and $s$ quarks. Such terms break the chiral symmetry explicitly and define corrections to the heavy-flavor ground-state masses.  We recall such terms in the chiral Lagrangian 
\begin{eqnarray}
&&\mathcal{L}_{\chi}=-\big( 4\,c_0-2\,c_1\big)\, H \,\bar{H}  \,{\tr} \chi_+ -2\,c_1\,H \,\chi _+\,\bar{H}
\nonumber\\
&& \quad \;\;+\,\big(2\,\tilde{c}_0-\tilde{c}_1\big)\,H^{\mu \nu }\,\bar{H}_{\mu \nu }\,{\tr}\chi _+
 +\tilde{c}_1\,H^{\mu \nu }\,\chi _+\,\bar{H}_{\mu \nu } \,,
\nonumber\\ \nonumber\\
&& \chi_\pm = {\textstyle \frac{1}{2}} \left(
e^{+i\,\frac{\Phi}{2\,f}} \,\chi_0 \,e^{+i\,\frac{\Phi}{2\,f}}
\pm e^{-i\,\frac{\Phi}{2\,f}} \,\chi_0 \,e^{-i\,\frac{\Phi}{2\,f}}
\right),
\label{def-chi}
\end{eqnarray}
where the quark-mass dependence is embodied in the diagonal matrix $ \chi_0 = 2\,B_0\, {\rm diag} (m_u,m_d,m_s)$, with the low-energy constant $B_0$. Throughout this work we assume perfect isospin symmetry with $m_u=m_d = m$. Those parameters contribute to the $J^P= 0^-$ and $J^P= 1^-$ meson masses at NLO
\begin{widetext}
\begin{eqnarray}
&& M_H^2= 
	\left\{
	\begin{aligned}
		& \Big(\bar M - \frac{3}{4}\Delta \Big)^2 + (4\,c_0 - 2\,c_1) \,\Pi_H^{(2),0} + 2\,c_1 \Pi_H^{(2),1} + \Pi_H^{\rm HO}\qquad {\rm if} \qquad H\in [ 0^- ]\\
		& \Big(\bar M + \frac{1}{4}\Delta \Big)^2 + (4\,\tilde c_0 - 2\,\tilde c_1)\, \Pi_H^{(2),0} + 2\,\tilde c_1 \Pi_H^{(2),1} + \Pi_H^{\rm HO}\qquad {\rm if} \qquad H\in [1^-]
	\end{aligned}
	\right. \,,
\label{H-Mass_ChPT-NLO} 
\end{eqnarray}
\end{widetext}
where the LECs $c_i = c_i(M_Q)$ and $\tilde c_i = \tilde c_i(M_Q)$ depend  on the heavy-quark mass $M_Q$. Throughout this work we will use an superscript $(c)$ and $(b)$ with e.g.
\begin{eqnarray}
c_i^{(c)} \equiv c_i (M_Q = M_c) \,,\qquad \tilde c_i^{(b)} \equiv \tilde c_i (M_Q = M_b)\,, 
\end{eqnarray}
for the open-charm and open-beauty systems, respectively. 
The chiral terms $\Pi_H^{(2),i}$ are linear combinations of the light-quark masses
\begin{eqnarray}
&& \Pi_{H}^{(2),0} = 2\,B_0(2\,m + m_s) \,, 
\label{tab:Pi}\\
&&  \Pi_{H}^{(2),1} = \left\{
	\begin{array}{ll} 2\,B_0 \,m  \quad \!& {\rm if} \!\!\quad H \in \{D,\,\, D^*, B,\,\, B^* \}\\ 
	                2\,B_0\,m_s \quad  \!&{\rm if} \! \!\quad H \in \{ D_s, D_s^*,B_s, B_s^* \}
\end{array}
	\right. \,. \nonumber
\end{eqnarray}
that contribute to chiral order $Q_\chi^2$, with 
\begin{align}
	Q_\chi\sim \sqrt{B_0m_q},\qquad q=u,d,s .
\end{align}
The higher-order chiral corrections $\Pi_H^{\rm HO}$ start from $O(Q_\chi^3)$ at the one-loop level.

The parameters $c_i$ and $\tilde c_i$  scale with the heavy-quark mass $M_Q\sim \bar M$ in the heavy-quark mass limit \cite{Wise:1992hn,Goity:1992tp}. 
It is useful to make this more explicit. We factor out their heavy-quark mass-independent part, denoted by  $C_i \sim \Lambda_\chi^{-1}$.
They are of dimension $-1$ with $\Lambda_\chi$ the chiral symmetry-breaking scale.
Higher-order corrections account for the heavy-quark symmetry breaking effects. 
At order $1/\bar M$, those effects enter. They consist of an overall shift to $C_i$ and a hyperfine splitting between the $c_i$ and $\tilde c_i$ \cite{Boyd:1994pa,Brambilla:2017hcq}. 
We introduce dimensionless parameters $\zeta_i$ and $\eta_i$ responsible for these two kinds of effects, and arrive at the following representation 
\begin{eqnarray}
	&&c_i(M_Q) = \bar M(M_Q) \Big(C_i + \frac{\zeta_i}{\bar M(M_Q)} - \frac{3}{4} \frac{\eta_i(M_Q)}{\bar M(M_Q)} \Big),\nonumber\\
	&&\tilde c_i(M_Q) = \bar M(M_Q) \Big(C_i + \frac{\zeta_i}{\bar M(M_Q)} +\frac{1}{4} \frac{\eta_i(M_Q)}{\bar M(M_Q)} \Big)\,.  \label{LEC-NLO_HQE}
\end{eqnarray}

The parameters $\eta_i = \eta_i(M_Q)$ depend on $M_Q$. The remaining four parameters $C_i$, $\zeta_i$ turn out to be independent on $M_Q$. This  will be seen in the next section, in which an explicit matching with HQET is worked out. 

In application of the LECs $c_i$ and $\tilde c_i$ in the charm sector, the $C_i$ can be determined modulo an unknown $\zeta_i$-dependence
\begin{align}
	C_i = \frac{1}{4\bar M^{(c)}} \Big(c_i^{(c)} +3\,\tilde c_i^{(c)} -4\, \zeta_i \Big)\,,
	\label{res-C_i} 
\end{align}
where we apply our notation $c^{(c)}_i\equiv c_i(M_c)$.
Similarly, the value of $\eta_i$ at $M_Q = M_c$ can be easily derived as 
\begin{align}
	\eta_i^{(c)} = \big( \tilde c_i^{(c)} - c_i^{(c)}\big) \,,
	\label{res-eta_i}
\end{align}
with again $\eta^{(c)}_i\equiv \eta_i(M_c)$.

In this work, the one-loop result is employed for the higher-order chiral correction terms $\Pi_H^{\rm HO}$. 
At the one-loop level, they receive contributions from  bubble and tadpole diagrams and their corresponding counterterms
\begin{align}
	\Pi_H^{\rm HO} = \Pi_H^{\rm bubble} + \Pi_H^{\rm tadpole} + \Pi_H^{\rm CT} \,.
	\label{def-PiHO}
\end{align}
For explicit expressions, Refs. \cite{Guo:2018kno,Guo:2018gyd} are referred to.
While more LECs are involved, they have been determined in the charm sector in Ref \cite{Guo:2018kno}. 
Using the leading-order scaling behavior, $\Pi_H^{\rm HO}\sim \bar M(M_Q)$, the LECs in the bottom sector can be well estimated.

\begin{widetext}
\section{Matching the chiral Lagrangian and the HQET}

So far we have not yet fully specified the LEC  $c^{(b)}_{0,1}$ and $\tilde c^{(b)}_{0,1}$ in \eqref{LEC-NLO_HQE}. While at leading order in the $1/M_Q$ expansion their values may be inferred from $c_{0,1}$ and $\tilde c_{0,1}$ at $M_Q = M_c$ this is no longer true at subleading order. Since such counterterms contribute at NLO in the 
chiral counting scheme it appears reasonable to consider the effect of order $1/M_Q$. 
Note that our approach considers chiral N$^2$LO effects in the heavy pseudoscalar and vector-meson masses. We now set up a more detailed matching with HQET, in which 
the heavy-meson masses take the form \cite{Falk:1992wt}
\begin{align}
	M_H(M_Q) =
	\left\{
	\begin{aligned}
	&M_Q + \bar \Lambda_{(H)} +\frac{\mu^2_{\pi(H)}}{2\,M_Q} -\frac{\mu_{G(H)}^2}{2\,M_Q} \qquad {\rm if} \qquad H\in [ 0^- ] \\
	&M_Q + \bar \Lambda_{(H)}+\frac{\mu^2_{\pi(H)}}{2\,M_Q} +\frac{\mu_{G(H)}^2}{6\,M_Q}
	\qquad {\rm if} \qquad H\in [ 1^- ]
	\end{aligned}
	\right. \,.
 \label{H-Mass_HQE}
\end{align}
\end{widetext}
The quantities $\bar \Lambda$ and $\mu^2_{\pi}$, $\mu^2_{G}$ come with an explicit index $(H)$,  that resolves the specifics of the light-quark content.
The $\bar \Lambda$ is the contribution from light degrees of freedom, and therefore $M_Q$ independent\cite{Manohar:2000dt}. 
The $\mu_\pi^2$ term accounts for the kinetic energy of the heavy quark in the meson's rest frame.
Due to reparametrization invariance, it is $M_Q$ independent as well \cite{Luke:1992cs}.
Finally, the $\mu_G^2$ is a chromomagnetic moment which leads to a hyperfine splitting between the  $0^-$ and $1^-$ B-mesons.
It depends on the heavy-quark mass.  
We assume that $\mu_G^2$ can be factorized as a product of the high-energy and low-energy contributions,
\begin{align}
	\mu_{G(H)}^2 = \hat C_{\rm cm}(M_Q)\, \hat \mu_{G(H)}^2\, ,
\end{align}
where the factor $\hat \mu_G$ accounts for low-energy contributions.
The high-energy contributions are incorporated in the renormalization-group\,(RG) invariant  Wilson coefficient $\hat C_{\rm cm}(M_Q)$\cite{Neubert:1993mb}. 
The RG evolution starts at a scale close to the heavy-quark mass $\mu \sim M_Q$, where the value of the Wilson coefficient is determined by matching the chromomagnetic moment from HQET to the multiloop calculations from the QCD Lagrangian. For our purposes it suffices to know the ratio $\hat C_{\rm cm}(M_b)/\hat C_{\rm cm}(M_c)$.
This ratio has been derived at the one-loop and two-loop level in \cite{Falk:1990pz} and \cite{Amoros:1997rx,Czarnecki:1997dz} respectively. The averaged result is 
\begin{align}
	R \equiv \frac{\hat C_{\rm cm}(M_b)}{\hat C_{\rm cm}(M_c)} \simeq 0.80(4) \,,
	\label{def-R}
\end{align}
where the uncertainty is estimated by the difference of the one-loop and the two-loop results. In the latest calculation, a poor convergence pattern has been claimed at the three-loop level \cite{Grozin:2007fh}. Therefore we refrain from using the three-loop result here.

The expansion moments $\bar \Lambda$, $\mu_\pi^2$, and $\hat \mu_G$ depend on  physical scales significantly lighter than $M_Q$: the light-quark masses $m_q$ and an intrinsic nonperturbative QCD scale $\Lambda$. While $\Lambda$ was once commonly regarded as the Landau scale $\Lambda_{\rm QCD}$ in literature \cite{Isgur:1989vq,Falk:1990pz,Kitazawa:1993bk}, we follow the way of Refs. \cite{Grinstein:1993ys,Boyd:1994pa} and identify $\Lambda\sim \Lambda_\chi$. It has been demonstrated in \cite{Randall:1992ww} that only via such an assignment, ChPT and HQET can be matched convincingly at the loop level.
Moreover, the size of $\Lambda$ should be comparable to the mass difference $\bar M - M_Q$ \cite{Falk:1990pz,Boyd:1994pa}. 
Using numerical values, this mass difference is $\sim 0.8$ GeV for both charm and bottom systems, indeed consistent with estimates of $\Lambda_\chi$.

In the chiral limit, the $\bar \Lambda$ scales as $\sim\Lambda$ whereas the $\mu_\pi^2$ and $\hat\mu_G^2$ scale as $\sim \Lambda^2$. 
We can expand the components $\bar \Lambda$, $\mu_\pi^2$ and $\hat\mu_G^2$ in powers of the light-quark masses around their chiral limit.
The corrections are suppressed by powers of $\sqrt{B_0 m_q}/\Lambda_\chi$.
The small-scale expansion scheme entails the scaling behavior $\Delta\sim \hat \mu_G^2/\bar M\sim Q_\chi$, and therefore the expansion parameters in the chiral and heavy-quark expansions are comparable
\begin{align}
	\frac{\sqrt{B_0 m_q}}{\Lambda_\chi}\sim \frac{\Lambda}{M_Q}\,.
\end{align}

By matching the mass formula \eqref{H-Mass_HQE} with the chiral result \eqref{H-Mass_ChPT-NLO}, we can recover the relations between the $\bar M, \Delta$ and the heavy-quark moments \cite{Wise:1992hn,Brambilla:2017hcq}. 
It is emphasized,  that the chiral structure of \eqref{H-Mass_ChPT-NLO} restricts the structure of the $O(Q_{\chi}^2)$ corrections to the heavy-quark expansion moments. From the matching we obtain that, 
$C_i$ and $\zeta_i$ are involved in the $O(Q_{\chi}^2)$ corrections of $\bar\Lambda$ and $\mu_\pi^2$ respectively. And they are indeed heavy-quark mass independent. In addition $\eta_i$ contributes to the $O(Q_{\chi}^2)$ corrections of $\mu_G^2$, and its scaling behavior is proportional to $\hat C_{\rm cm}$. We summarize,
\begin{align}
	\frac{\bar M^{(b)}\Delta^{(b)}}{\bar M^{(c)}\Delta^{(c)}} = 
	\frac{\eta_i^{(b)}}{\eta_i^{(c)}} = R \,.
	\label{res-R}
\end{align}
Using (\ref{res-R}) with $R \simeq 0.80$ together with \eqref{res-eta_i} we are left with three unknown parameters $\bar M^{(b)}$ and $ \zeta_{0,1}$, given the charm-sector LECs $\bar M^{(c)}$, $\Delta^{(c)}$, $c_i^{(c)}, \tilde c_i^{(c)}$ as inputs. The $M_Q$ independent parameters $C_i$ follow from \eqref{res-C_i}. 

The three unknown parameters $\bar M^{(b)}$ and $ \zeta_{0,1}$ are determined by a fit with our chiral mass formula to the empirical values of the four $B$-meson ground-state masses. Here we admit a residual systematic uncertainty of $ 5$ MeV in the heavy-meson masses. Such a value was used in our previous open-charm system studies \cite{Guo:2018kno}. It reflects the accuracy level at which we expect our one-loop chiral formula to hold. 
The results of $\bar M^{(b)}$, $ \zeta_{0,1}$ are shown in Tab. \ref{tab:Fits}. In this table, we also show the parameters involved in the expansion \eqref{LEC-NLO_HQE} together  with the associated LECs.
In the fits, we used the results of \cite{Guo:2018kno}, for the inputs of the LECs in the charm sector.
In the charm sector, four sets of fitted results are determined according to the lattice data on $D$-meson ground-state masses and $\pi D$ s-wave scattering process. They are named as Fits 1-4.
We will recall some of the fitting details in the following discussion.

\begin{table}[t]
\setlength{\tabcolsep}{2.1mm}
\renewcommand{\arraystretch}{1.2}
\hspace{-0.1cm}
\begin{tabular}{c|r|r|r|r}
\hline
                     &{ Fit 1 }          & {  Fit 2}          & {  Fit 3}   & {  Fit 4}      \\ 
 \hline
$\bar M^{(b)}$[GeV]  &$5.3743$         
                     &$4.8540$                
                     &$5.3303$         
                     &$5.3666$         \\ 
$\zeta_0$            &$0.0921$         
                     &$-1.5072$                          
                     &$-0.0839$        
                     &$0.0523$         \\
$\zeta_1$            &$0.1689$         
                     &$0.1233$                          
                     &$0.1585$         
                     &$0.1678$         \\
                     \hline
$C_0[{\rm GeV}^{-1}]$&$0.0602$
                     &$0.8777$  
                     &$0.1774$  
                     &$0.1145$  \\
$C_1[{\rm GeV}^{-1}]$&$0.2376$
                     &$0.3916$  
                     &$0.3382$  
                     &$0.3445$ \\
$\eta_0^{(b)}$       &$-0.0145$
                     &$-0.0302$  
                     &$-0.0176$  
                     &$-0.0170$  \\
$\eta_1^{(b)}$       &$-0.0238$
                     &$0.0318$  
                     &$-0.0276$  
                     &$-0.0238$    \\   [2pt]  \hline                      
$\Delta^{(b)}$[GeV]        &$0.0562$         &$0.0643$         &$0.0563$         &$0.0568$          \\ \hline                     
$c_0^{(b)}$                &$0.4262$         
                           &$2.7757$      
                           &$0.8750$        
                           &$0.6797$    \\ 
$\tilde c_0^{(b)}$         &$0.4117$        
                           &$2.7455$        
                           &$0.8574$         
                           &$0.6628$    \\  
$c_1^{(b)}$                &$1.4637$         
                           &$2.0002$        
                           &$1.9820$        
                           &$2.0345$   \\ 
$\tilde c_1^{(b)}$         &$1.4399$         
                           &$2.0320$       
                           &$1.9544$       
                           &$2.0107$     \\   [2pt] \hline
$\chi^2/N$           &$0.94  $         
                     &$0.10  $                          
                     &$0.89  $         
                     &$0.92  $         \\[2pt]                     
\hline
\end{tabular}
\caption{The low-energy parameters in \eqref{LEC-NLO_HQE}, corresponding to Fits 1-4 in the charm sector \cite{Guo:2018kno}. The $\chi^2/N$ is the chi-square per data point, with the number of data points $N = 4$ and an ad hoc systematic error estimate of $ 5$ MeV. }
\label{tab:Fits}
\end{table}

Consider first the scenarios 
of Fit 1,\,3,\,4. The masses of the $B$-meson ground states can be reproduced within the systematic error of 5\,MeV. All of the 3 fits give modest heavy-quark corrections to the leading-order expectations of $c_{0,1}$ and $\tilde c_{0,1}$. 
At leading order, $c_{0,1}$ and $\tilde c_{0,1}$ are about 2.5 times larger at $M_Q = M_b$ as compared to their values at $M_Q = M_c$. For convenience we recall the ranges $c_0\sim \tilde c_0\sim (0.2-0.3)$ and $c_1\sim \tilde c_1\sim (0.6-0.9)$ at $M_Q = M_c$ from \cite{Guo:2018kno}. 

Scenarios 3 and 4 show quite similar values for the LEC. This is not the case for scenario 1. Here we recall a decisive distinction. Both Fit 1 and Fit 2 
did not consider QCD lattice data on the $\pi D$ s-wave scattering 
process \cite{Moir:2016srx}. While Fit 1, nevertheless, appears reasonably consistent with the $\pi D$ phase shift and inelasticity parameters as given in \cite{Moir:2016srx}, this is not the case for Fit 2. The key feature of Fit 3 and also Fit 4 is their compatibility with the lattice data on the $\eta D$ phase shift. Such data play a crucial role in the determination of the LEC. Based on this observation we would disfavor scenarios 1 and 2. In this context it is amusing to observe that Fit 2 should be rejected also based on an unnaturally large value of the $\zeta_0$ parameter as shown in Tab. \ref{tab:Fits}. This is so despite the fact that it comes with the best chi-square value for the reproduction of the B-meson masses. The corresponding $c_0 \sim \tilde c_0 \sim 2.7$ at $M_Q = M_b$, are nearly ten times larger than their charmed counterparts $c_0 \sim \tilde c_0 \sim 0.3$. This implied a serious violation of the leading-order scaling behavior. Such  large LECs lead to unnaturally large higher order corrections. Therefore, altogether we exclude Fit 2 from our further analysis.

\section{Scatterings with coupled-channel dynamics}

The chiral Lagrangian predicts the formation of $J^P = 0^+$ and $J^P = 1^+$ resonance state 
as a consequence of coupled-channel final-state interactions of the Goldstone bosons with 
the ground-state heavy mesons with  $J^P = 0^-$ and $J^P = 1^-$ quantum numbers. 
Such states are an unavoidable consequence  of the flavor SU(3) chiral Lagrangian. 
We focus on the resonances generated by the $s$-wave scatterings with open-beauty quantum numbers. Here the leading order coupled-channel interaction is predicted by the Tomozawa-Weinberg theorem in terms of the "known" parameter $f$. 

A resonance is dynamically generated from a scattering process when the reaction amplitude  
contains a pole in the complex $s$ plane. 
The scattering amplitude with manifest $s$-channel unitarity is obtained from a self-consistent summation 
\begin{eqnarray}
	T_{ab}(s) =  V_{ab}(s) + \sum_{c,\,d}V_{ac}(s)\,J_{cd}(s)\, T_{db}(s)\,,
	\label{BSE}
\end{eqnarray}
with given out- and in-going two-body states $a$ and $b$.
The $T$-matrix exhibits poles in the complex $s$ plane that we can determine by extending the definition of $T(s)$ into the higher Riemann sheets of the $s$ plane.
The potential $V_{ab}(s)$ is obtained with an on-shell condition \cite{Lutz:2001yb}, and set equivalent to the scattering amplitude as derived from the chiral Lagrangian at the matching point  $\sqrt{s}=\mu_M$. 
It receives tree-level chiral symmetry-breaking contributions from the Lagrangian \eqref{def-chi}. Additional terms are implied by the LEC that imply  
the one-loop $\Pi_H^{\rm HO}$ structures decomposed in (\ref{def-PiHO}). 
The latter will also contribute to $V_{ab}(s)$ as tree-level chiral symmetry-preserving corrections, see Ref \cite{Guo:2018kno}.
Following \cite{Kolomeitsev:2003ac}, we set $\mu_M = M_{B^{(*)}}$ for $S=0,2$ and $\mu_M=M_{B_s^{(*)}}$ for $S=\pm 1$ scatterings with the total quantum numbers $J^P = 0^+ (1^+)$.
The diagonal analytic matrix $J(s)$ function is universal as it leads to 
a scattering amplitude $T(s)$ that is consistent with the coupled-channel unitarity condition and the microcausality condition. Such an approach can be justified 
if short-range forces largely dominate the system. Along the real axis from each threshold, there is a branch cut defining the doorway for the higher Riemann sheets.
For an $n$-dimensional coupled-channel system there are $2^n$ Riemann sheets, and
we use the signature $(\pm,\cdots,\pm)$ as introduced in \cite{Guo:2018gyd} to label a specific one.

\begin{table}[t]
\setlength{\tabcolsep}{3.8mm}
\renewcommand{\arraystretch}{1.4}
\centering
\begin{tabular}{c|l|c}
\hline
      & $(I,S)= (1/2,0)$                             & $(I,S)=  (0,1)$        \\ \hline
\multicolumn{3}{l}{$J^P = 0^+$} \\[2pt] \hline
Fit 1 & $5.5202^{+282}_{-209}-0.0923^{+463}_{-254}\,i$   & $5.6296^{+431}_{-395}$ \\
Fit 3 & $5.5137^{+179}_{-225}-0.1073^{+623}_{-387}\,i$ & $5.5689^{+614}_{-544}$ \\
Fit 4 & $5.5126^{+154}_{-196}-0.1120^{+595}_{-384}\,i$ & $5.5755^{+610}_{-535}$ \\
TW    & $5.5207_{-190}^{+260}-0.0905^{+456}_{-243}\,i$ & $5.6495^{+353}_{-310}$ \\ \hline
\multicolumn{3}{l}{$J^P = 1^+$} \\[2pt] \hline
Fit 1 & $5.5652^{+283}_{-208}-0.0915^{+460}_{-252}\,i$   & $5.6763^{+428}_{-391}$ \\   
Fit 3 & $5.5595^{+179}_{-224}-0.1071^{+624}_{-384}\,i$ & $5.6179^{+610}_{-542}$ \\   
Fit 4 & $5.5586^{+156}_{-198}-0.1115^{+597}_{-382}\,i$ & $5.6242^{+605}_{-533}$ \\   
TW    & $5.5658_{-190}^{+260}-0.0903^{+455}_{-242}\,i$ & $5.6959^{+354}_{-311}$ \\
\hline
\end{tabular}
\caption{Complex pole masses (in GeV) of the flavor antitriplet states with $J^P = 0^+$ and $1^+$. The relevant Riemann sheets are $(-,+,+)$ and $(+,+)$ for $(I,S) = (1/2,0)$ and $(0,1)$.}
\label{tab:pole3}
\end{table}

We start with a discussion of the flavor antitriplet channels. 
Poles in the complex $s$-plane are found from the parameter sets 1, 3 and 4. The complex pole masses are compared with the results from the Tomozawa-Weinberg interaction in Tab.\,\ref{tab:pole3}. 
To estimate the theoretical error, we allow a deviation of the matching points from their natural values for $|\Delta \mu_M| = 0.1$ GeV. 
We first look at the states with isospin and strangeness $(I,S)=(0,1)$.
A pole below the $BK$ threshold is found on the physical Riemann sheet in the $J^P = 0^+$ scattering amplitude always. It is the open-beauty partner of $D_{s0}^*(2317)$\,. 
The pole mass is $5.59(8)$\,GeV.
Comparing to the leading-order result at $5.65(3)$\,GeV, the higher-order chiral corrections slightly reduce the pole mass. This result is somewhat lower than previous  predictions with values above 5.7 GeV\,\cite{Colangelo:2012xi,Altenbuchinger:2013vwa,Lang:2015hza,Du:2017zvv}. In the axial-vector sector, the open-beauty partner of $D_{s1}(2460)$ is found as a bound state at $5.64(8)$\,GeV, 
which should be compared with previous predictions at above 5.75\,GeV\,\cite{Colangelo:2012xi,Altenbuchinger:2013vwa,Lang:2015hza,Du:2017zvv}.
Besides the bound states in the $(I,S) = (0,1)$ channels, broad resonances are found in the $(I,S) = (1/2,0)$ channels with poles in the unphysical Riemann sheet denoted by $(-,+,+)$. Their broad charmed partners were extensively discussed in previous theoretical studies\,\cite{Kolomeitsev:2003ac,Hofmann:2003je,Lutz:2007sk,Altenbuchinger:2013vwa,Yao:2015qia,Du:2017zvv}. Our prediction of their pole masses are $\big(5.52(3) -0.12(5)\,i\big)$\,GeV 
for the $0^+$ state and $\big(5.57(3) -0.12(5)\,i\big)$\,GeV
for the $1^+$ state. 
Both of them are in agreement with previous theoretical predictions \cite{Kolomeitsev:2003ac,Guo:2006rp,Altenbuchinger:2013vwa,Du:2017zvv}.

\begin{table}[t]
\setlength{\tabcolsep}{1.5mm}
\renewcommand{\arraystretch}{1.4}
\centering
\begin{tabular}{c|l|c}
\hline
      & $(I,S)= (1/2,0)$                             & $(I,S)=  (1,1)$        \\ \hline
\multicolumn{3}{l}{$J^P = 0^+$} \\[2pt] \hline
Fit 1 & $5.8104^{+095}_{-099}-0.0148^{-023}_{+010}\,i$ & $5.8045^{+261}_{-205}-0.0730^{+127}_{-064}\,i$ \\
Fit\,3 & $5.7274^{+258}_{-144}-0.0595^{-230}_{+158}\, i$ & $5.7945^{+147}_{-154}-0.1871^{+266}_{-219} \,i$ \\
Fit\,4 & $5.7422^{+285}_{-171}-0.0517^{-233}_{+186} \,i$ & $5.7868^{+133}_{-134}-0.1938^{+234}_{-190} \,i$ \\
TW    & $5.7730^{+148}_{-149}-0.0201^{-22}_{+7}\, i$ & $5.7900^{+319}_{-245}-0.0719^{+107}_{-52} i$ \\ \hline
\multicolumn{3}{l}{$J^P = 1^+$} \\[2pt] \hline
Fit 1 & $5.8592^{+091}_{-096}-0.0153^{-025}_{+011}\,i$ & $5.8558^{+264}_{-208}-0.0730^{+127}_{-063}\,i$ \\   
Fit\,3 & $5.7747^{+254}_{-141}-0.0638^{-237}_{+160}\,i$ & $5.8406^{+146}_{-150}-0.1913^{+256}_{-211}\,i$ \\   
Fit\,4 & $5.7894^{+279}_{-166}-0.0564^{-241}_{+186}\,i$ & $5.8333^{+133}_{-132}-0.1972^{+227}_{-184}\,i$ \\   
TW    & $5.8192^{+148}_{-150}-0.0198^{-22}_{+7}\,i$ &  $5.8362^{+320}_{-246}-0.0708^{+105}_{-51}i$ \\
\hline
\end{tabular}
\caption{Complex pole masses (in GeV) of two flavor sextet states with $J^P = 0^+$ and $1^+$. The relevant Riemann sheets are $(-,+,+)$ and $(-,+)$ for $(I,S) = (1/2,0)$ and $(1,1)$.}
\label{tab:pole6}
\end{table}
 
In the charmed and beauty sector, there are further poles belonging to a flavor sextet. In Tab. \ref{tab:pole6}, we listed the complex pole masses found in the sextet channels with $(I,S) = (1/2,0)$ and $(1,1)$ from Fits 1,3, and 4. The other sextet channel, with $(I,S) = (0,-1)$, shows a pole in the vicinity of the scattering threshold within the range of our theoretical uncertainty always. The resonances with $(I,S) = (1/2,0)$ were also obtained in previous works \cite{Guo:2006rp,Du:2017zvv} with masses $(5.84-5.85)$ GeV and $(5.88-5.91)$ GeV, respectively, for $J^P = 0^+$ and $1^+$. Our predicted values are significantly smaller than those. The $J^P = 0^+$ member with $(I,S) = (1,1)$ is of particular interest. It has been speculated that the controversial $X(5568)$ state has such quantum numbers (see e.g. \cite{Chen:2016spr,Brambilla:2019esw}). 
In  \cite{Kolomeitsev:2003ac}, it was shown that the LO Tomozawa-Weinberg interaction implies the existence of an exotic state with $J^P = 0^+$ and $(I,S) = (1,1)$ quantum numbers at $\sqrt{s} \simeq\big(5.79(3)-0.07(1)\,i\big)$\,GeV. We demonstrate that this prediction is quite stable against higher-order corrections from our Fits 1,3, and 4. The 
pole mass comes at $\big(5.80(3) - 0.14(8)\,i\big)$\,GeV as shown in Tab. \ref{tab:pole6}.    
We conclude that the $X(5568)$ cannot be a chiral excitation, i.e. it is not explained convincingly by chiral coupled-channel dynamics. This supports previous such claims  \cite{Albaladejo:2016eps,Sun:2016tmz}.  

\section{Summary}

We studied open-beauty mesons with $J^P = 0^+$ and $1^+$ quantum numbers. Such states were predicted as a consequence of coupled-channel interactions based on the chiral SU(3) Lagrangian. Already the leading order Tomozawa-Weinberg interaction implies attractive forces in the flavor antitriplet and sextet channels between the Goldstone bosons and the 
heavy-meson ground states with  $J^P = 0^-$ and $1^-$ quantum numbers.  

In this article the role of the next-to-leading order chiral interactions in the s-wave open-beauty meson scattering processes was scrutinized.   
The LECs are derived mainly from corresponding  LECs  as obtained previously from global fits to the QCD lattice dataset in the charm sector \cite{Guo:2018kno}. Where possible additional direct data form the beauty sector were taken into account. The heavy-quark scaling behavior is constrained by the  RG-invariant Wilson coefficient for the chromomagnetic moment. We employ the results of the Wilson coefficient calculated at the 2-loop level.

We find that in the antitriplet but also in the exotic flavor sextet,
the chiral correction terms lead to minor effects in the 
$J^P = 0^+$ and $1^+$ pole masses only. This confirms the semiquantitative predictions 
made almost two decades ago by one of the authors. Our refined values should be used in ongoing experimental searches and QCD lattice simulations. It is noted, however, that like in the open-charm sector, we expect the light-quark mass dependence in the flavor antitriplet and sextet states to be significant. This  should be investigated further, in particular for the $\pi B$ s-wave phase shifts.

\vskip-0.3cm
\section{Acknowledgements}

Xiang-Dong Gao and Daniel Mohler are acknowledged for stimulating discussions.


\bibliographystyle{elsarticle-num}
\bibliography{literature}

\begin{thebibliography}{10}
\expandafter\ifx\csname url\endcsname\relax
  \def\url#1{\texttt{#1}}\fi
\expandafter\ifx\csname urlprefix\endcsname\relax\def\urlprefix{URL }\fi
\expandafter\ifx\csname href\endcsname\relax
  \def\href#1#2{#2} \def\path#1{#1}\fi

\bibitem{Lutz:2001yb}
M.~F.~M. Lutz, E.~E. Kolomeitsev, {Relativistic chiral SU(3) symmetry, large
  N(c) sum rules and meson baryon scattering}, Nucl. Phys. A700 (2002)
  193--308.
\newblock \href {http://arxiv.org/abs/nucl-th/0105042}
  {\path{arXiv:nucl-th/0105042}}, \href
  {http://dx.doi.org/10.1016/S0375-9474(01)01312-4}
  {\path{doi:10.1016/S0375-9474(01)01312-4}}.

\bibitem{DiPierro:2001dwf}
M.~Di~Pierro, E.~Eichten, {Excited Heavy - Light Systems and Hadronic
  Transitions}, Phys. Rev. D 64 (2001) 114004.
\newblock \href {http://arxiv.org/abs/hep-ph/0104208}
  {\path{arXiv:hep-ph/0104208}}, \href
  {http://dx.doi.org/10.1103/PhysRevD.64.114004}
  {\path{doi:10.1103/PhysRevD.64.114004}}.

\bibitem{Albaladejo:2018mhb}
M.~Albaladejo, P.~Fernandez-Soler, J.~Nieves, P.~G. Ortega, {Contribution of
  constituent quark model $c\bar{s}$ states to the dynamics of the
  $D_{s0}^*(2317)$ and $D_{s1}(2460)$ resonances}, Eur. Phys. J. C 78~(9)
  (2018) 722.
\newblock \href {http://arxiv.org/abs/1805.07104} {\path{arXiv:1805.07104}},
  \href {http://dx.doi.org/10.1140/epjc/s10052-018-6176-3}
  {\path{doi:10.1140/epjc/s10052-018-6176-3}}.

\bibitem{Kolomeitsev:2003ac}
E.~Kolomeitsev, M.~F.~M.~Lutz, {On Heavy light meson resonances and chiral
  symmetry}, Phys. Lett. B582 (2004) 39--48.
\newblock \href {http://arxiv.org/abs/hep-ph/0307133}
  {\path{arXiv:hep-ph/0307133}}, \href
  {http://dx.doi.org/10.1016/j.physletb.2003.10.118}
  {\path{doi:10.1016/j.physletb.2003.10.118}}.

\bibitem{Guo:2006fu}
F.-K. Guo, P.-N. Shen, H.-C. Chiang, R.-G. Ping, B.-S. Zou, {Dynamically
  generated 0+ heavy mesons in a heavy chiral unitary approach}, Phys. Lett. B
  641 (2006) 278--285.
\newblock \href {http://arxiv.org/abs/hep-ph/0603072}
  {\path{arXiv:hep-ph/0603072}}, \href
  {http://dx.doi.org/10.1016/j.physletb.2006.08.064}
  {\path{doi:10.1016/j.physletb.2006.08.064}}.

\bibitem{Lang:2014yfa}
C.~Lang, L.~Leskovec, D.~Mohler, S.~Prelovsek, R.~Woloshyn, {Ds mesons with DK
  and D*K scattering near threshold}, Phys. Rev. D 90~(3) (2014) 034510.
\newblock \href {http://arxiv.org/abs/1403.8103} {\path{arXiv:1403.8103}},
  \href {http://dx.doi.org/10.1103/PhysRevD.90.034510}
  {\path{doi:10.1103/PhysRevD.90.034510}}.

\bibitem{Lutz:2007sk}
M.~F.~M.~Lutz, M.~Soyeur, {Radiative and isospin-violating decays of
  D(s)-mesons in the hadrogenesis conjecture}, Nucl.Phys. A813 (2008) 14--95.
\newblock \href {http://arxiv.org/abs/0710.1545} {\path{arXiv:0710.1545}},
  \href {http://dx.doi.org/10.1016/j.nuclphysa.2008.09.003}
  {\path{doi:10.1016/j.nuclphysa.2008.09.003}}.

\bibitem{Guo:2008gp}
F.-K. Guo, C.~Hanhart, S.~Krewald, U.-G. Mei{\ss}ner, {Subleading contributions
  to the width of the D*(s0)(2317)}, Phys. Lett. B 666 (2008) 251--255.
\newblock \href {http://arxiv.org/abs/0806.3374} {\path{arXiv:0806.3374}},
  \href {http://dx.doi.org/10.1016/j.physletb.2008.07.060}
  {\path{doi:10.1016/j.physletb.2008.07.060}}.

\bibitem{Liu:2012zya}
L.~Liu, K.~Orginos, F.-K. Guo, C.~Hanhart, U.-G. Mei{\ss}ner, {Interactions of
  charmed mesons with light pseudoscalar mesons from lattice QCD and
  implications on the nature of the $D_{s0}^*(2317)$}, Phys. Rev. D 87~(1)
  (2013) 014508.
\newblock \href {http://arxiv.org/abs/1208.4535} {\path{arXiv:1208.4535}},
  \href {http://dx.doi.org/10.1103/PhysRevD.87.014508}
  {\path{doi:10.1103/PhysRevD.87.014508}}.

\bibitem{Altenbuchinger:2013vwa}
M.~Altenbuchinger, L.~S. Geng, W.~Weise, {Scattering lengths of Nambu-Goldstone
  bosons off $D$ mesons and dynamically generated heavy-light mesons}, Phys.
  Rev. D 89~(1) (2014) 014026.
\newblock \href {http://arxiv.org/abs/1309.4743} {\path{arXiv:1309.4743}},
  \href {http://dx.doi.org/10.1103/PhysRevD.89.014026}
  {\path{doi:10.1103/PhysRevD.89.014026}}.

\bibitem{Guo:2018tjx}
Z.-H. Guo, L.~Liu, U.-G. Mei\ss{}ner, J.~A. Oller, A.~Rusetsky, {Towards a
  precise determination of the scattering amplitudes of the charmed and
  light-flavor pseudoscalar mesons}, Eur. Phys. J. C 79~(1) (2019) 13.
\newblock \href {http://arxiv.org/abs/1811.05585} {\path{arXiv:1811.05585}},
  \href {http://dx.doi.org/10.1140/epjc/s10052-018-6518-1}
  {\path{doi:10.1140/epjc/s10052-018-6518-1}}.

\bibitem{Guo:2018kno}
X.-Y. Guo, Y.~Heo, M.~F.~M. Lutz, {On chiral extrapolations of charmed meson
  masses and coupled-channel reaction dynamics}, Phys. Rev. D98~(1) (2018)
  014510.
\newblock \href {http://arxiv.org/abs/1801.10122} {\path{arXiv:1801.10122}},
  \href {http://dx.doi.org/10.1103/PhysRevD.98.014510}
  {\path{doi:10.1103/PhysRevD.98.014510}}.

\bibitem{Guo:2018gyd}
X.-Y. Guo, Y.~Heo, M.~F.~M. Lutz, {On chiral excitations with exotic quantum
  numbers}, Phys. Lett. B 791 (2019) 86--91.
\newblock \href {http://arxiv.org/abs/1809.01311} {\path{arXiv:1809.01311}},
  \href {http://dx.doi.org/10.1016/j.physletb.2019.02.022}
  {\path{doi:10.1016/j.physletb.2019.02.022}}.

\bibitem{Gayer:2021xzv}
L.~Gayer, N.~Lang, S.~M. Ryan, D.~Tims, C.~E. Thomas, D.~J. Wilson,
  {Isospin-1/2 $D\pi$ scattering and the lightest $D_0^\ast$ resonance from
  lattice QCD}\href {http://arxiv.org/abs/2102.04973}
  {\path{arXiv:2102.04973}}.

\bibitem{Hofmann:2003je}
J.~Hofmann, M.~F.~M.~Lutz, {Open charm meson resonances with negative
  strangeness}, Nucl.Phys. A733 (2004) 142--152.
\newblock \href {http://arxiv.org/abs/hep-ph/0308263}
  {\path{arXiv:hep-ph/0308263}}, \href
  {http://dx.doi.org/10.1016/j.nuclphysa.2003.12.013}
  {\path{doi:10.1016/j.nuclphysa.2003.12.013}}.

\bibitem{D0:2016mwd}
V.~Abazov, et~al., {Evidence for a $B_s^0 \pi^\pm$ state}, Phys. Rev. Lett.
  117~(2) (2016) 022003.
\newblock \href {http://arxiv.org/abs/1602.07588} {\path{arXiv:1602.07588}},
  \href {http://dx.doi.org/10.1103/PhysRevLett.117.022003}
  {\path{doi:10.1103/PhysRevLett.117.022003}}.

\bibitem{Aaij:2016iev}
R.~Aaij, et~al., {Search for Structure in the $B_s^0\pi^\pm$ Invariant Mass
  Spectrum}, Phys. Rev. Lett. 117~(15) (2016) 152003, [Addendum: Phys.Rev.Lett.
  118, 109904 (2017)].
\newblock \href {http://arxiv.org/abs/1608.00435} {\path{arXiv:1608.00435}},
  \href {http://dx.doi.org/10.1103/PhysRevLett.117.152003}
  {\path{doi:10.1103/PhysRevLett.117.152003}}.

\bibitem{Aaltonen:2017voc}
T.~Aaltonen, et~al., {A search for the exotic meson $X(5568)$ with the Collider
  Detector at Fermilab}, Phys. Rev. Lett. 120~(20) (2018) 202006.
\newblock \href {http://arxiv.org/abs/1712.09620} {\path{arXiv:1712.09620}},
  \href {http://dx.doi.org/10.1103/PhysRevLett.120.202006}
  {\path{doi:10.1103/PhysRevLett.120.202006}}.

\bibitem{Sirunyan:2017ofq}
A.~M. Sirunyan, et~al., {Search for the X(5568) state decaying into
  $\mathrm{B}^{0}_{\mathrm{s}}\pi^{\pm}$ in proton-proton collisions at
  $\sqrt{s} = $ 8 TeV}, Phys. Rev. Lett. 120~(20) (2018) 202005.
\newblock \href {http://arxiv.org/abs/1712.06144} {\path{arXiv:1712.06144}},
  \href {http://dx.doi.org/10.1103/PhysRevLett.120.202005}
  {\path{doi:10.1103/PhysRevLett.120.202005}}.

\bibitem{Aaboud:2018hgx}
M.~Aaboud, et~al., {Search for a Structure in the $B^0_s \pi^\pm$ Invariant
  Mass Spectrum with the ATLAS Experiment}, Phys. Rev. Lett. 120~(20) (2018)
  202007.
\newblock \href {http://arxiv.org/abs/1802.01840} {\path{arXiv:1802.01840}},
  \href {http://dx.doi.org/10.1103/PhysRevLett.120.202007}
  {\path{doi:10.1103/PhysRevLett.120.202007}}.

\bibitem{Agaev:2016mjb}
S.~Agaev, K.~Azizi, H.~Sundu, {Mass and decay constant of the newly observed
  exotic $X(5568)$ state}, Phys. Rev. D 93~(7) (2016) 074024.
\newblock \href {http://arxiv.org/abs/1602.08642} {\path{arXiv:1602.08642}},
  \href {http://dx.doi.org/10.1103/PhysRevD.93.074024}
  {\path{doi:10.1103/PhysRevD.93.074024}}.

\bibitem{Lang:2016jpk}
C.~Lang, D.~Mohler, S.~Prelovsek, {$B_s\pi^+$ scattering and search for X(5568)
  with lattice QCD}, Phys. Rev. D 94 (2016) 074509.
\newblock \href {http://arxiv.org/abs/1607.03185} {\path{arXiv:1607.03185}},
  \href {http://dx.doi.org/10.1103/PhysRevD.94.074509}
  {\path{doi:10.1103/PhysRevD.94.074509}}.

\bibitem{Burns:2016gvy}
T.~Burns, E.~Swanson, {Interpreting the X (5568)}, Phys. Lett. B 760 (2016)
  627--633.
\newblock \href {http://arxiv.org/abs/1603.04366} {\path{arXiv:1603.04366}},
  \href {http://dx.doi.org/10.1016/j.physletb.2016.07.049}
  {\path{doi:10.1016/j.physletb.2016.07.049}}.

\bibitem{Lu:2016kxm}
J.-X. Lu, X.-L. Ren, L.-S. Geng, {$B_s\pi $ -- $B\bar{K}$ interactions in
  finite volume and $X(5568)$}, Eur. Phys. J. C 77~(2) (2017) 94.
\newblock \href {http://arxiv.org/abs/1607.06327} {\path{arXiv:1607.06327}},
  \href {http://dx.doi.org/10.1140/epjc/s10052-017-4660-9}
  {\path{doi:10.1140/epjc/s10052-017-4660-9}}.

\bibitem{Chen:2016mqt}
W.~Chen, H.-X. Chen, X.~Liu, T.~Steele, S.-L. Zhu, {Decoding the $X(5568)$ as a
  fully open-flavor $su\bar b\bar d$ tetraquark state}, Phys. Rev. Lett.
  117~(2) (2016) 022002.
\newblock \href {http://arxiv.org/abs/1602.08916} {\path{arXiv:1602.08916}},
  \href {http://dx.doi.org/10.1103/PhysRevLett.117.022002}
  {\path{doi:10.1103/PhysRevLett.117.022002}}.

\bibitem{Liu:2016xly}
X.-H. Liu, G.~Li, {Could the observation of X(5568) be a result of the near
  threshold rescattering effects?}, Eur. Phys. J. C 76~(8) (2016) 455.
\newblock \href {http://arxiv.org/abs/1603.00708} {\path{arXiv:1603.00708}},
  \href {http://dx.doi.org/10.1140/epjc/s10052-016-4308-1}
  {\path{doi:10.1140/epjc/s10052-016-4308-1}}.

\bibitem{Yang:2016sws}
Z.~Yang, Q.~Wang, U.-G. Mei{\ss}ner, {Where does the X(5568) structure come
  from?}, Phys. Lett. B 767 (2017) 470--473.
\newblock \href {http://arxiv.org/abs/1609.08807} {\path{arXiv:1609.08807}},
  \href {http://dx.doi.org/10.1016/j.physletb.2017.01.023}
  {\path{doi:10.1016/j.physletb.2017.01.023}}.

\bibitem{Ke:2018stp}
H.-W. Ke, X.-Q. Li, {How can $X^{\pm}(5568)$ escape detection?}, Phys. Lett. B
  785 (2018) 301--303.
\newblock \href {http://arxiv.org/abs/1802.08823} {\path{arXiv:1802.08823}},
  \href {http://dx.doi.org/10.1016/j.physletb.2018.07.071}
  {\path{doi:10.1016/j.physletb.2018.07.071}}.

\bibitem{Wise:1992hn}
M.~B. Wise, {Chiral perturbation theory for hadrons containing a heavy quark},
  Phys. Rev. D45~(7) (1992) R2188.
\newblock \href {http://dx.doi.org/10.1103/PhysRevD.45.R2188}
  {\path{doi:10.1103/PhysRevD.45.R2188}}.

\bibitem{Goity:1992tp}
J.~L. Goity, {Chiral perturbation theory for SU(3) breaking in heavy meson
  systems}, Phys. Rev. D46 (1992) 3929--3936.
\newblock \href {http://arxiv.org/abs/hep-ph/9206230}
  {\path{arXiv:hep-ph/9206230}}, \href
  {http://dx.doi.org/10.1103/PhysRevD.46.3929}
  {\path{doi:10.1103/PhysRevD.46.3929}}.

\bibitem{Boyd:1994pa}
C.~G. Boyd, B.~Grinstein, {Chiral and heavy quark symmetry violation in B
  decays}, Nucl. Phys. B442 (1995) 205--227.
\newblock \href {http://arxiv.org/abs/hep-ph/9402340}
  {\path{arXiv:hep-ph/9402340}}, \href
  {http://dx.doi.org/10.1016/S0550-3213(95)00005-4}
  {\path{doi:10.1016/S0550-3213(95)00005-4}}.

\bibitem{Brambilla:2017hcq}
N.~Brambilla, J.~Komijani, A.~Kronfeld, A.~Vairo, {Relations between
  Heavy-light Meson and Quark Masses}, Phys. Rev. D 97~(3) (2018) 034503.
\newblock \href {http://arxiv.org/abs/1712.04983} {\path{arXiv:1712.04983}},
  \href {http://dx.doi.org/10.1103/PhysRevD.97.034503}
  {\path{doi:10.1103/PhysRevD.97.034503}}.

\bibitem{Falk:1992wt}
A.~F. Falk, M.~Neubert, {Second order power corrections in the heavy quark
  effective theory. 1. Formalism and meson form-factors}, Phys. Rev. D47 (1993)
  2965--2981.
\newblock \href {http://arxiv.org/abs/hep-ph/9209268}
  {\path{arXiv:hep-ph/9209268}}, \href
  {http://dx.doi.org/10.1103/PhysRevD.47.2965}
  {\path{doi:10.1103/PhysRevD.47.2965}}.

\bibitem{Manohar:2000dt}
A.~Manohar, M.~Wise, \href{http://books.google.de/books?id=codDQK5OQDIC}{Heavy
  Quark Physics}, Cambridge Monographs on Particle Physics, Nuclear Physics and
  Cosmology, Cambridge University Press, 2007.
\newline\urlprefix\url{http://books.google.de/books?id=codDQK5OQDIC}

\bibitem{Luke:1992cs}
M.~E. Luke, A.~V. Manohar, {Reparametrization invariance constraints on heavy
  particle effective field theories}, Phys. Lett. B286 (1992) 348--354.
\newblock \href {http://arxiv.org/abs/hep-ph/9205228}
  {\path{arXiv:hep-ph/9205228}}, \href
  {http://dx.doi.org/10.1016/0370-2693(92)91786-9}
  {\path{doi:10.1016/0370-2693(92)91786-9}}.

\bibitem{Neubert:1993mb}
M.~Neubert, {Heavy quark symmetry}, Phys. Rept. 245 (1994) 259--396.
\newblock \href {http://arxiv.org/abs/hep-ph/9306320}
  {\path{arXiv:hep-ph/9306320}}, \href
  {http://dx.doi.org/10.1016/0370-1573(94)90091-4}
  {\path{doi:10.1016/0370-1573(94)90091-4}}.

\bibitem{Falk:1990pz}
A.~F. Falk, B.~Grinstein, M.~E. Luke, {Leading mass corrections to the heavy
  quark effective theory}, Nucl. Phys. B357 (1991) 185--207.
\newblock \href {http://dx.doi.org/10.1016/0550-3213(91)90464-9}
  {\path{doi:10.1016/0550-3213(91)90464-9}}.

\bibitem{Amoros:1997rx}
G.~Amoros, M.~Beneke, M.~Neubert, {Two loop anomalous dimension of the
  chromomagnetic moment of a heavy quark}, Phys. Lett. B 401 (1997) 81--90.
\newblock \href {http://arxiv.org/abs/hep-ph/9701375}
  {\path{arXiv:hep-ph/9701375}}, \href
  {http://dx.doi.org/10.1016/S0370-2693(97)00345-6}
  {\path{doi:10.1016/S0370-2693(97)00345-6}}.

\bibitem{Czarnecki:1997dz}
A.~Czarnecki, A.~G. Grozin, {HQET chromomagnetic interaction at two loops},
  Phys. Lett. B 405 (1997) 142--149, [Erratum: Phys.Lett.B 650, 447 (2007)].
\newblock \href {http://arxiv.org/abs/hep-ph/9701415}
  {\path{arXiv:hep-ph/9701415}}, \href
  {http://dx.doi.org/10.1016/S0370-2693(97)00587-X}
  {\path{doi:10.1016/S0370-2693(97)00587-X}}.

\bibitem{Grozin:2007fh}
A.~G. Grozin, P.~Marquard, J.~H. Piclum, M.~Steinhauser, {Three-Loop
  Chromomagnetic Interaction in HQET}, Nucl. Phys. B 789 (2008) 277--293.
\newblock \href {http://arxiv.org/abs/0707.1388} {\path{arXiv:0707.1388}},
  \href {http://dx.doi.org/10.1016/j.nuclphysb.2007.08.012}
  {\path{doi:10.1016/j.nuclphysb.2007.08.012}}.

\bibitem{Isgur:1989vq}
N.~Isgur, M.~B. Wise, {Weak Decays of Heavy Mesons in the Static Quark
  Approximation}, Phys. Lett. B232 (1989) 113--117.
\newblock \href {http://dx.doi.org/10.1016/0370-2693(89)90566-2}
  {\path{doi:10.1016/0370-2693(89)90566-2}}.

\bibitem{Kitazawa:1993bk}
N.~Kitazawa, T.~Kurimoto, {Heavy meson effective theory with 1/M(Q)
  correction}, Phys. Lett. B323 (1994) 65--70.
\newblock \href {http://arxiv.org/abs/hep-ph/9312225}
  {\path{arXiv:hep-ph/9312225}}, \href
  {http://dx.doi.org/10.1016/0370-2693(94)00047-6}
  {\path{doi:10.1016/0370-2693(94)00047-6}}.

\bibitem{Grinstein:1993ys}
B.~Grinstein, {On a Precise Calculation of (f(B(s)) / f(B)) / (f(D(s)) / f(D))
  and Its Implications on the Interpretation of B Anti-B Mixing}, Phys. Rev.
  Lett. 71 (1993) 3067--3069.
\newblock \href {http://arxiv.org/abs/hep-ph/9308226}
  {\path{arXiv:hep-ph/9308226}}, \href
  {http://dx.doi.org/10.1103/PhysRevLett.71.3067}
  {\path{doi:10.1103/PhysRevLett.71.3067}}.

\bibitem{Randall:1992ww}
L.~Randall, E.~Sather, {The QCD scale in the heavy quark expansion}, Phys. Rev.
  D49 (1994) 6236--6239.
\newblock \href {http://arxiv.org/abs/hep-ph/9211268}
  {\path{arXiv:hep-ph/9211268}}, \href
  {http://dx.doi.org/10.1103/PhysRevD.49.6236}
  {\path{doi:10.1103/PhysRevD.49.6236}}.

\bibitem{Moir:2016srx}
G.~Moir, M.~Peardon, S.~M. Ryan, C.~E. Thomas, D.~J. Wilson, {Coupled-Channel
  $D\pi$, $D\eta$ and $D_{s}\bar{K}$ Scattering from Lattice QCD}, JHEP 10
  (2016) 011.
\newblock \href {http://arxiv.org/abs/1607.07093} {\path{arXiv:1607.07093}},
  \href {http://dx.doi.org/10.1007/JHEP10(2016)011}
  {\path{doi:10.1007/JHEP10(2016)011}}.

\bibitem{Colangelo:2012xi}
P.~Colangelo, F.~De~Fazio, F.~Giannuzzi, S.~Nicotri, {New meson spectroscopy
  with open charm and beauty}, Phys. Rev. D 86 (2012) 054024.
\newblock \href {http://arxiv.org/abs/1207.6940} {\path{arXiv:1207.6940}},
  \href {http://dx.doi.org/10.1103/PhysRevD.86.054024}
  {\path{doi:10.1103/PhysRevD.86.054024}}.

\bibitem{Lang:2015hza}
C.~B. Lang, D.~Mohler, S.~Prelovsek, R.~M. Woloshyn, {Predicting positive
  parity B$_s$ mesons from lattice QCD}, Phys. Lett. B750 (2015) 17--21.
\newblock \href {http://arxiv.org/abs/1501.01646} {\path{arXiv:1501.01646}},
  \href {http://dx.doi.org/10.1016/j.physletb.2015.08.038}
  {\path{doi:10.1016/j.physletb.2015.08.038}}.

\bibitem{Du:2017zvv}
M.-L. Du, M.~Albaladejo, P.~Fernández-Soler, F.-K. Guo, C.~Hanhart, U.-G.
  Mei{\ss}ner, J.~Nieves, D.-L. Yao, {Towards a new paradigm for heavy-light
  meson spectroscopy}, Phys. Rev. D 98~(9) (2018) 094018.
\newblock \href {http://arxiv.org/abs/1712.07957} {\path{arXiv:1712.07957}},
  \href {http://dx.doi.org/10.1103/PhysRevD.98.094018}
  {\path{doi:10.1103/PhysRevD.98.094018}}.

\bibitem{Yao:2015qia}
D.-L. Yao, M.-L. Du, F.-K. Guo, U.-G. Mei{\ss}ner, {One-loop analysis of the
  interactions between charmed mesons and Goldstone bosons}, JHEP 11 (2015)
  058.
\newblock \href {http://arxiv.org/abs/1502.05981} {\path{arXiv:1502.05981}},
  \href {http://dx.doi.org/10.1007/JHEP11(2015)058}
  {\path{doi:10.1007/JHEP11(2015)058}}.

\bibitem{Guo:2006rp}
F.-K. Guo, P.-N. Shen, H.-C. Chiang, {Dynamically generated 1+ heavy mesons},
  Phys. Lett. B 647 (2007) 133--139.
\newblock \href {http://arxiv.org/abs/hep-ph/0610008}
  {\path{arXiv:hep-ph/0610008}}, \href
  {http://dx.doi.org/10.1016/j.physletb.2007.01.050}
  {\path{doi:10.1016/j.physletb.2007.01.050}}.

\bibitem{Chen:2016spr}
H.-X. Chen, W.~Chen, X.~Liu, Y.-R. Liu, S.-L. Zhu, {A review of the open charm
  and open bottom systems}, Rept. Prog. Phys. 80~(7) (2017) 076201.
\newblock \href {http://arxiv.org/abs/1609.08928} {\path{arXiv:1609.08928}},
  \href {http://dx.doi.org/10.1088/1361-6633/aa6420}
  {\path{doi:10.1088/1361-6633/aa6420}}.

\bibitem{Brambilla:2019esw}
N.~Brambilla, S.~Eidelman, C.~Hanhart, A.~Nefediev, C.-P. Shen, C.~E. Thomas,
  A.~Vairo, C.-Z. Yuan, {The $XYZ$ states: experimental and theoretical status
  and perspectives}\href {http://arxiv.org/abs/1907.07583}
  {\path{arXiv:1907.07583}}.

\bibitem{Albaladejo:2016eps}
M.~Albaladejo, J.~Nieves, E.~Oset, Z.-F. Sun, X.~Liu, {Can $X(5568)$ be
  described as a $B_s\pi$, $B\bar{K}$ resonant state?}, Phys. Lett. B 757
  (2016) 515--519.
\newblock \href {http://arxiv.org/abs/1603.09230} {\path{arXiv:1603.09230}},
  \href {http://dx.doi.org/10.1016/j.physletb.2016.04.033}
  {\path{doi:10.1016/j.physletb.2016.04.033}}.

\bibitem{Sun:2016tmz}
B.-X. Sun, F.-Y. Dong, J.-L. Pang, {Study of X(5568) in a unitary
  coupled-channel approximation of $B \bar{K}$ and $B_s \pi$}, Chin. Phys. C
  41~(7) (2017) 074104.
\newblock \href {http://arxiv.org/abs/1609.04068} {\path{arXiv:1609.04068}},
  \href {http://dx.doi.org/10.1088/1674-1137/41/7/074104}
  {\path{doi:10.1088/1674-1137/41/7/074104}}.

\end{thebibliography}

\end{document}